\documentclass{iau} 
\usepackage{graphicx}

\title[Next generation spectroscopic analysis] 
{Next generation spectroscopic analysis for large samples of massive stars}

\author[Joachim M.\ Bestenlehner]   
{Joachim M. Bestenlehner$^1$
}

\affiliation{$^1$Department of Physics \& Astronomy, Hounsfield Road, \\ University of Sheffield, S3 7RH, UK \\ email: {\tt j.m.bestenlehner@sheffield.ac.uk} 
}
\pubyear{2022}
\volume{361}  
\jname{Massive Stars Near and Far}
\editors{N. St-Louis, J. S. Vink \& J. Mackey, eds.}
\begin{document}

\maketitle

\begin{abstract}
Upcoming large-scale spectroscopic surveys such as WEAVE and 4MOST will provide thousands of spectra of massive stars, which need to be analysed in an efficient and homogeneous way. Studies on massive stars are usually based on samples of a few hundred objects which pushes current spectroscopic analysis tools to their limits because visual inspection is necessary to verify the spectroscopic fit.

The novel spectroscopic analysis pipeline takes advantage of the statistics that large samples provide, and determines the model error to account for imperfections in stellar atmosphere codes due to simplified, wrong or missing physics. Considering observational plus model uncertainties improve spectroscopic fits. The pipeline utilises the entire spectrum rather than selected diagnostic lines allowing a wider range of temperature from B to early O stars to be analysed. A small fraction of stars like peculiar, contaminated or spectroscopic binaries require visual inspection, which are identified through their larger uncertainties.



\end{abstract}

\firstsection 
\section{Introduction}

\begin{figure}[t]
\begin{center}
 \includegraphics[width=3.4in]{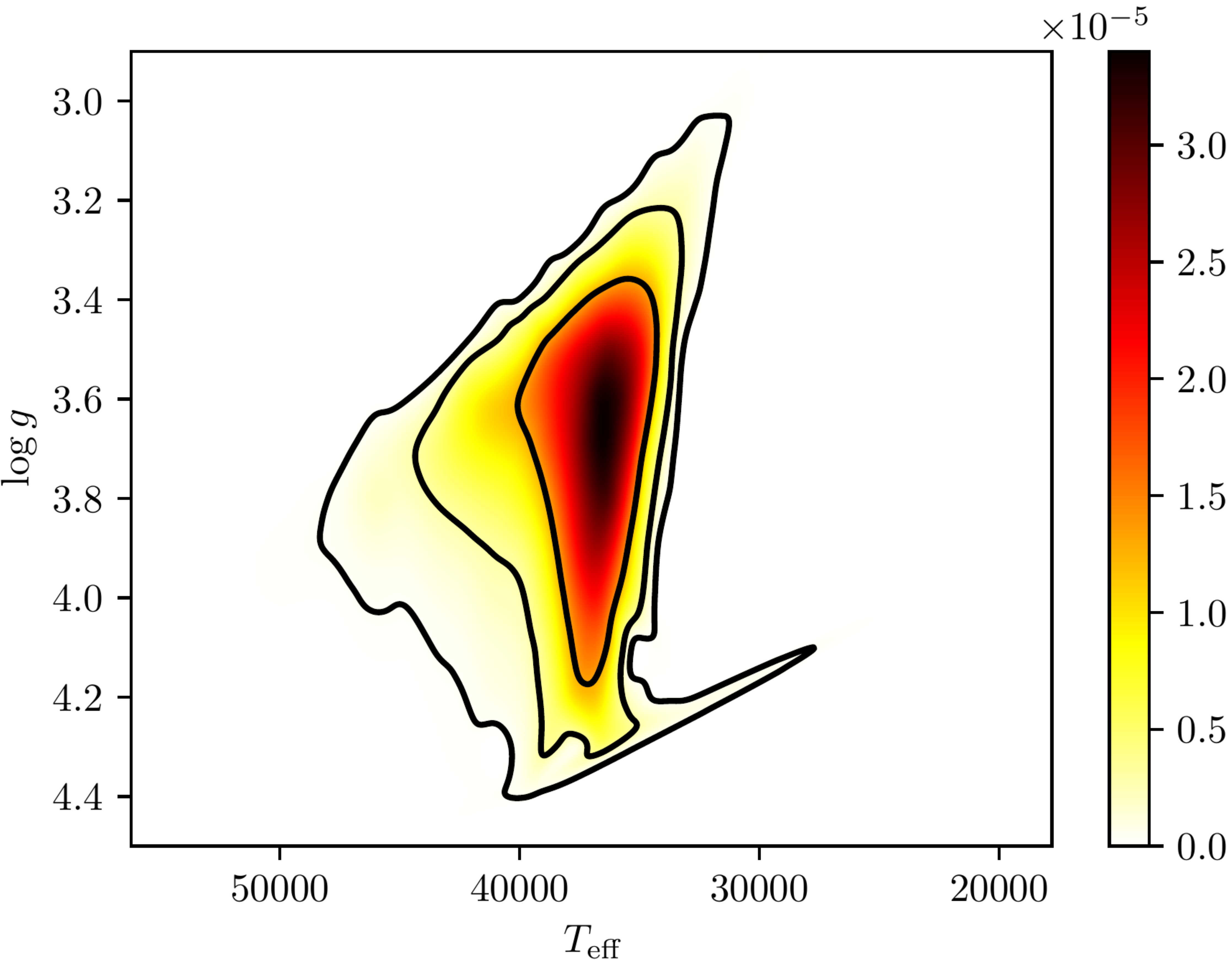} 
 \caption{Probability heat map of surface gravity vs. effective temperature. Contours indicate confidence intervals of 68\%, 95\% and 99.7\%.}
   \label{fig1}
\end{center}
\end{figure}

Historically and still today, the most common way to analyse massive stars is by ``eye'', which limits the sample size to 10s of stars. This means that stellar parameters as well as uncertainties are estimated rather than determined. Sample of a couple of hundreds of stars are usually analysed with a $\chi^2$-minimisation algorithm. Multi-dimensional probability distribution functions are obtained depending on the number of free parameters and uncertainties are then defined on confidence intervals rather than Gaussian standard deviations (Fig.~\ref{fig1}). Those uncertainties can be highly asymmetric and very large in the case of degenerated parameters. In the massive star community there are 2 main flavours of $\chi^2$-minimisation algorithms, grid based (e.g. \cite[Sim{\'o}n-D{\'\i}az et al. 2011]{SimonDiaz2011}; \cite[Castro et al. 2012]{Castro2012}; \cite[Bestenlehner et al. 2014]{Bestenlehner2014}) and Generic Algorithm on the basis of natural selection (e.g. \cite[Mokiem et al. 2007]{Mokiem2007}; \cite[Brands et al. 2022]{Brands2022}). All those pipelines use a pre-defined selection of spectral lines for their analysis. 

However, the drawback of the $\chi^2$-method is that parameter estimates can be strongly distorted by minimizing differences between synthetic spectra and real observations. This requires human verification of the spectroscopic fit, which is not feasible for large samples in excess of a couple of hundreds stars. In addition, synthetic spectra based on stellar atmosphere models are imperfect due to missing or simplified physics, insufficient atomic data, and so forth. Therefore, model uncertainties should also be budgeted into the parameter determination.

\section{Method}
The pipeline presented here is a grid-based $\chi^2$-minimisation algorithm, but uses information of the entire available spectral range instead of specific spectral lines. The observational error spectrum is included into the error matrix $\mathrm{N}$ and the $\chi^2$ has the following form
\begin{equation}
\chi^2 = (\vec{d} - \mathrm{R}\vec{s})^{\mathrm T}\mathrm{N}^{-1}(\vec{d} - \mathrm{R}\vec{s})
\end{equation}
with data $\vec{d}$, observed spectra, response matrix $\mathrm{R}$ and idealised synthetic model spectrum $\vec{s}$. Synthetic spectra are de-idealised by asking the reverse question: for a given model spectrum $\vec{s}$, which data $\vec{d}$ can reproduce the model. This is implemented by substituting $\vec{s}$ with $\vec{t}^{[p]} + \vec{u}$. $\vec{t}^{[p]}$ is the idealised model spectrum defined by stellar parameters $p$ and additional error term $\vec{u}$. The model error of a single star does not contain much information, but should be determined utilising the statistic a large sample of homogeneously observed stars provides.

To test and verify the methodology we used the VLT/MUSE observation of $\sim 250$ OB stars from \cite[Castro et al. (2018)]{Castro2018}. The VLT/MUSE data cover the wavelength range between 4600 to 9300 {\AA} at spectral resolution of 2000 to 4000. The spectrum normalisation is fully automated simulating the work-flow of the analysis of a large dataset.

The grid of synthetic spectra was computed with the non-LTE stellar atmosphere and radiative transfer code FASTWIND (\cite[Santolaya-Rey et al. 1997]{Santolaya1997}; \cite[Puls et al. 2005]{Puls2005}; \cite[Rivero Gonz{\'a}lez et al. 2011]{Rivero2011}). The VLT/MUSE data cover a less used wavelength range for hot star and we extended the line list to longer wavelength ($\sim$ 10\,000 {\AA}) beyond the tested range of FASTWIND (4000 to 7000 {\AA}). Around 150\,000 stellar atmosphere models were computed with varying stellar parameters and chemical composition of He and CNO. Of those models $\sim$20\% did not converged properly and were removed. Still the number was too large to individually check all stellar models and we trust that the majority of those remaining are okay. 

\section{Results}
\begin{figure}[t]
\begin{center}
 \includegraphics[width=0.495\textwidth]{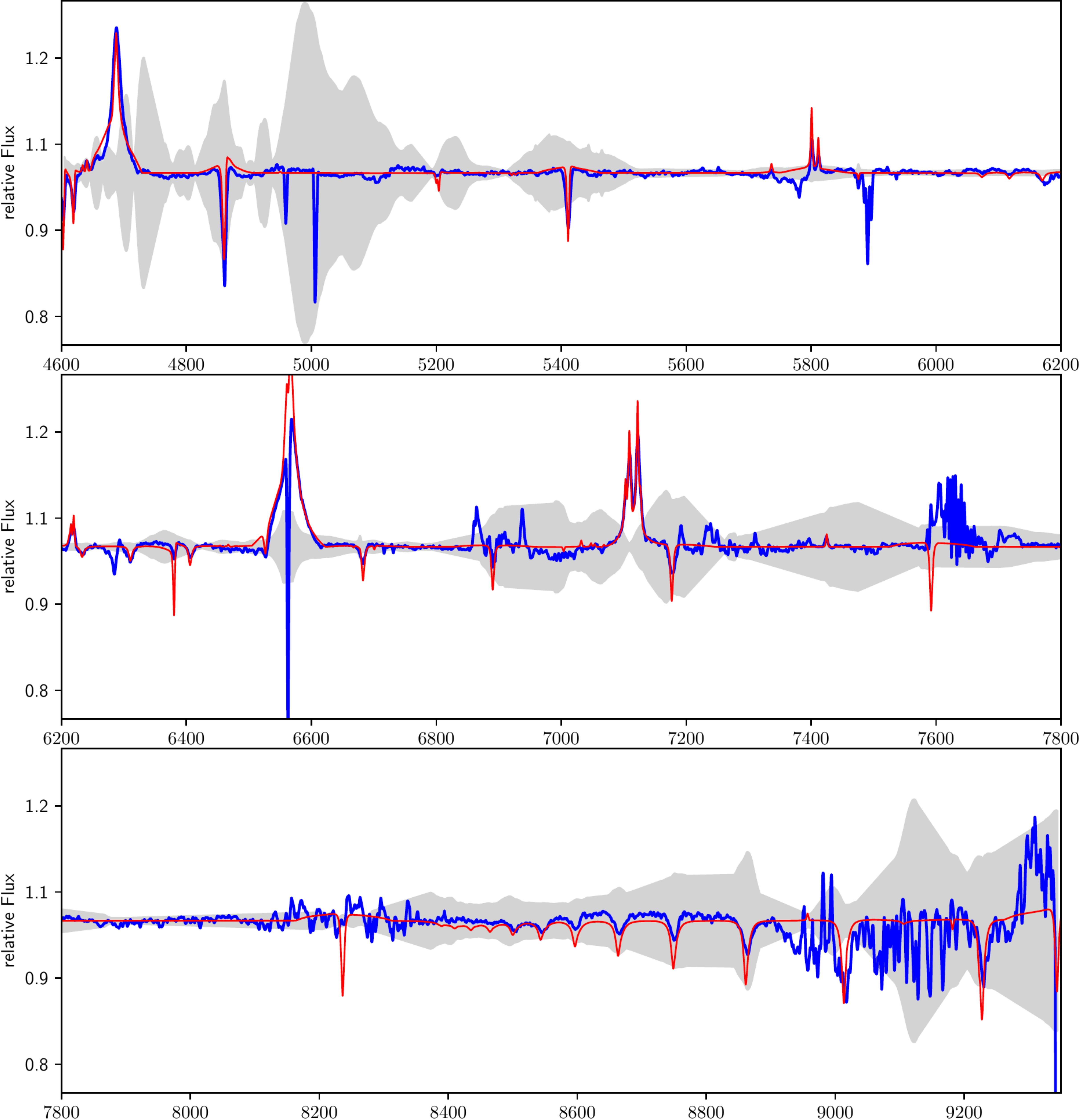} 
 \includegraphics[width=0.495\textwidth]{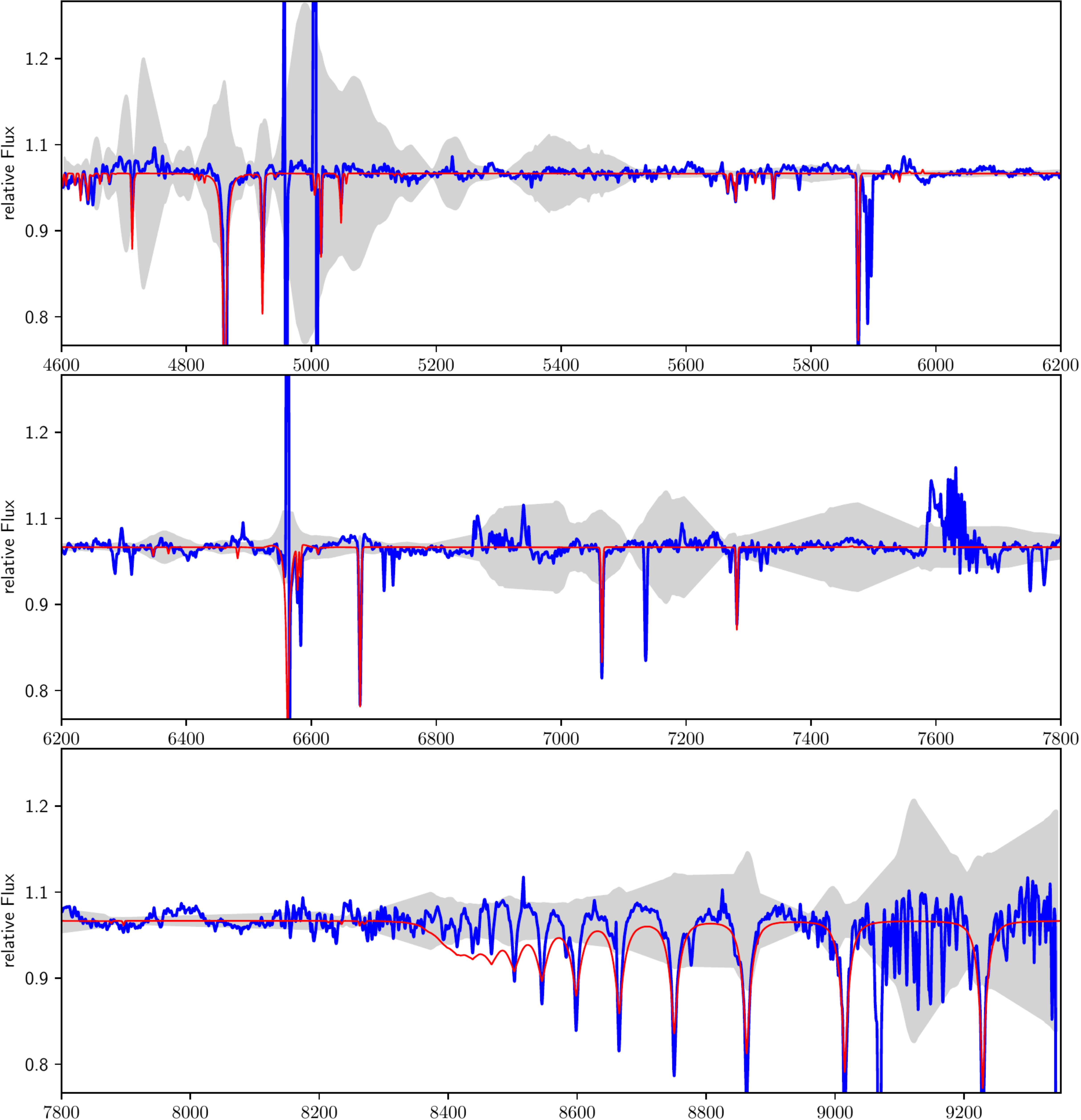}
 \caption{Spectroscopic fit of an early Of supergiant (left) and a early-mid B dwarf (right). Blue solid line is the observation, red solid line the synthetic spectrum and the grey shaded area is the square-root of the diagonal elements of the covariant-matrix calculated by the pipeline.}
   \label{fig2}
\end{center}
\end{figure}
The following parameters were derived: effective temperature $T_{\rm eff}$, surface gravity $\log g$, helium composition $Y$ and estimates on mass-loss rate $\dot{M}$ and CN chemical abundances. Figure~\ref{fig2} shows the spectroscopic fit of an Of supergiant and a B dwarf with $\Delta T_{\rm eff} \approx 25\,000$\,K and $\Delta \log\dot{M} \gtrsim 2.5$\,dex. This highlights that stars covering of large spectral type range can be successfully and reliably analysed with a single pipeline set up at the same time. However, issues occur for low signal to noise spectra (S/N $\lesssim 10$ to 15) and spectra with strong nebular lines.

\begin{figure}[t]
\begin{center}
 \includegraphics[width=0.495\textwidth]{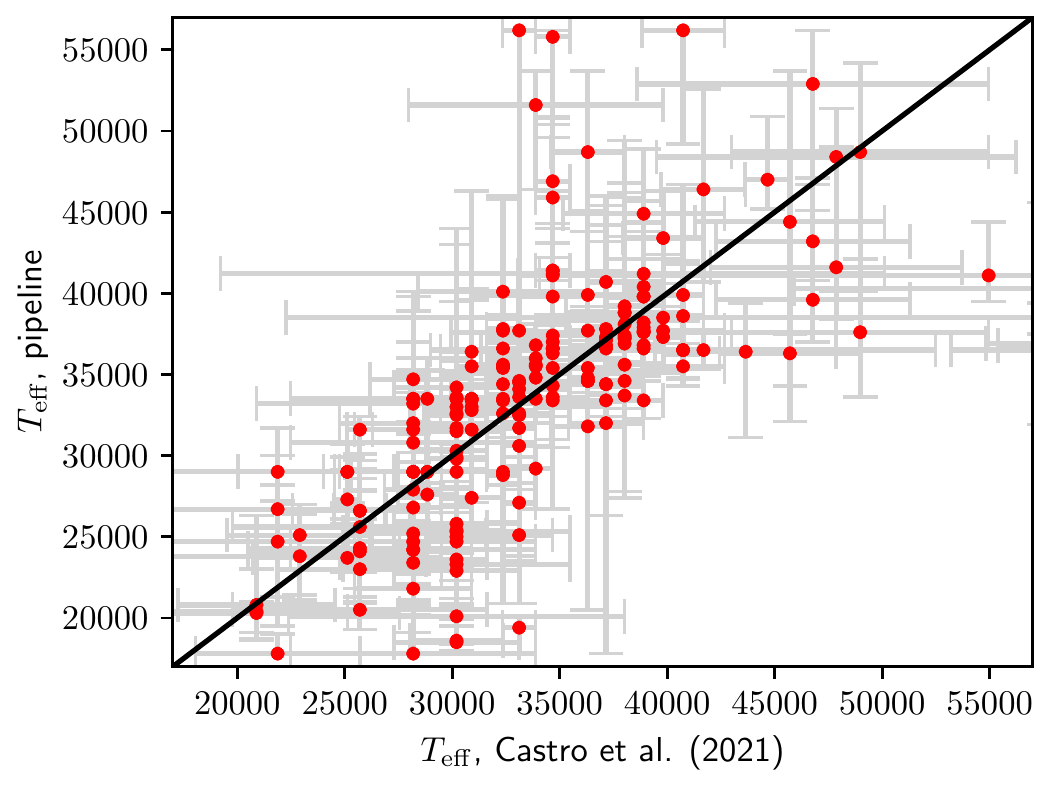} 
 \includegraphics[width=0.495\textwidth]{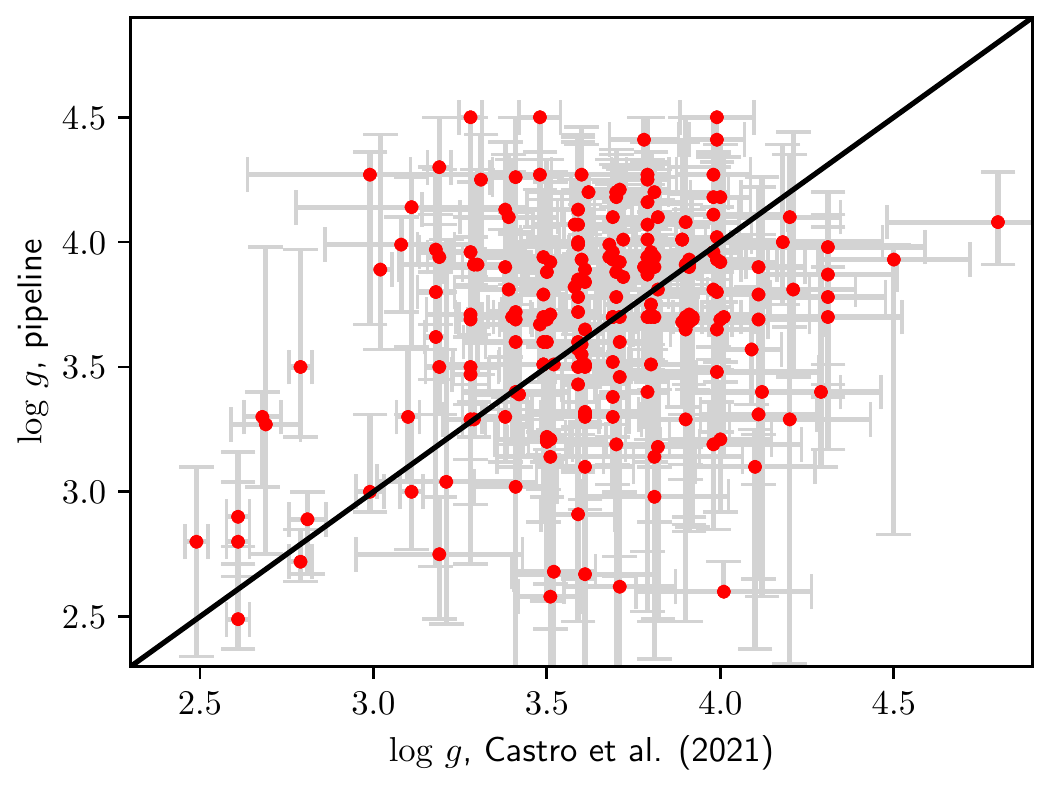} 
 \caption{Effective temperatures (left) and surface gravities (right) determined by the pipeline vs. the results from \cite[Castro et al. (2021)]{Castro2021}.}
   \label{fig3}
\end{center}
\end{figure}

In Fig.~\ref{fig3} we compare our results with the ones of \cite[Castro et al. (2021)]{Castro2021} which is based on the ionisation balance of selected HeI and HeII and the wings of $H_\mathrm{\beta}$. In contrast, we used all H, He plus CNO, Si and Mg metal lines available in the VLT/MUSE wavelength range. In general there is a good agreement for the effective temperature, but the pipeline predicted in correct temperatures for spectra with strong nebular contaminations. The picture is inconclusive for the surface gravity. Even though we also utilised the Paschen lines, $\log g$ is basically unconstrained in both studies.

\begin{figure}[t]
\begin{center}
 \includegraphics[width=3.4in]{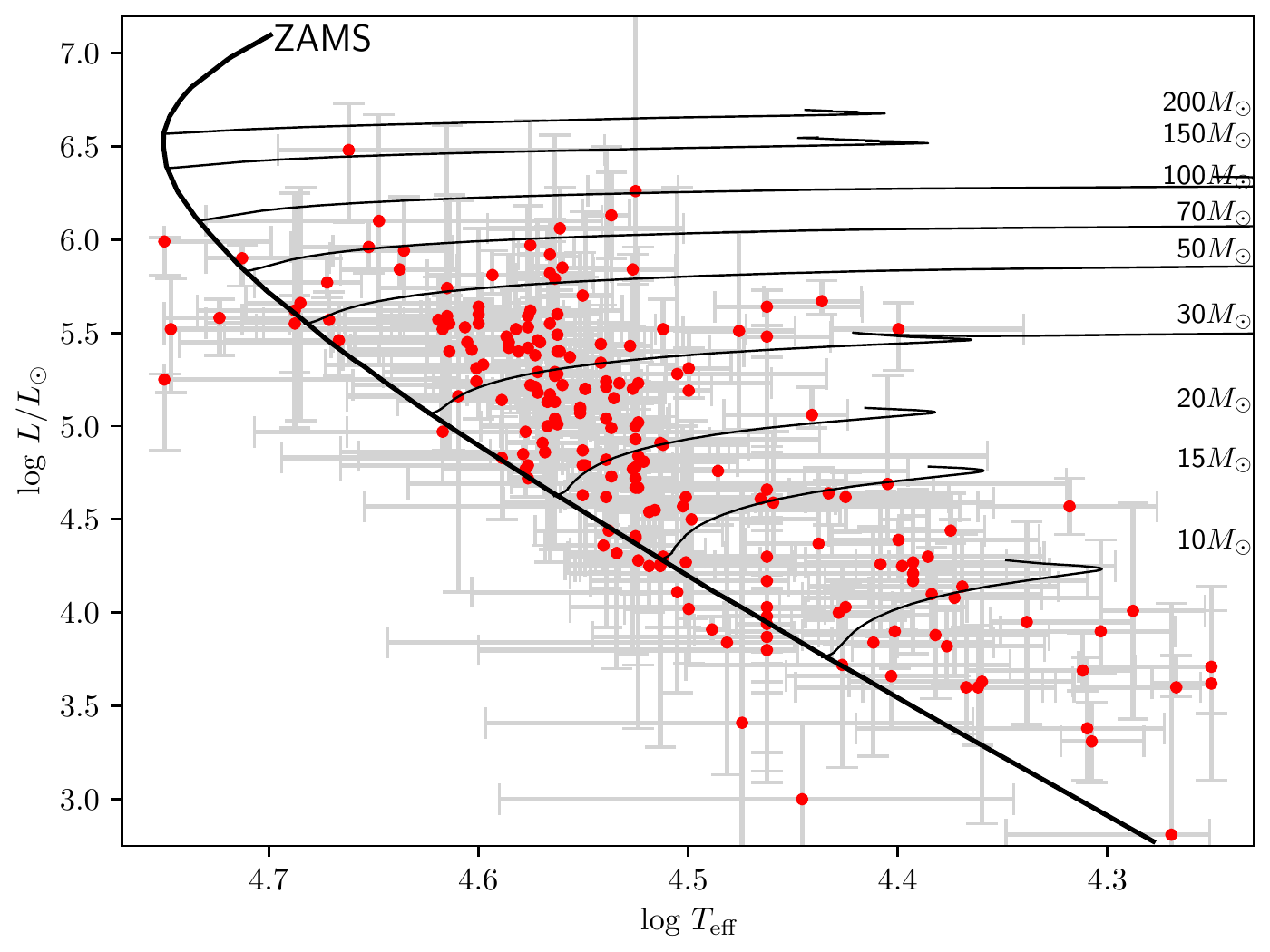} 
 \caption{Hertzsprung-Russell diagram of the analysed stars using the VLT/MUSE data from \cite[Castro et al. (2018)]{Castro2018}. Thin black lines are stellar evolutionary tracks by \cite[Brott et al. (2011)]{Brott2011} and \cite[K\"ohler et al. (2015)]{Koehler2015}.}
   \label{fig4}
\end{center}
\end{figure}

The Hertzprung-Russell diagram (HRD, Fig.~\ref{fig4}) shows that as expected most stars are populated near and to the cool side of the zero-age main-sequence (ZAMS). There are a couple of exceptions but their uncertainties do not exclude a cooler location in agreement with the majority of the sources. This can be improved by including a meaningful prior into the analysis, e.g. based on evolutionary tracks, and could increase the accuracy of the results, as we used only a flat prior. For example, only hydrogen deficient stars are found to be on the hot side of the ZAMS. A prior would give the star a higher probability to be found either on the hot or cool side of the ZAMS depending on its helium composition.

\begin{figure}[t]
\begin{center}
 \includegraphics[width=3.4in]{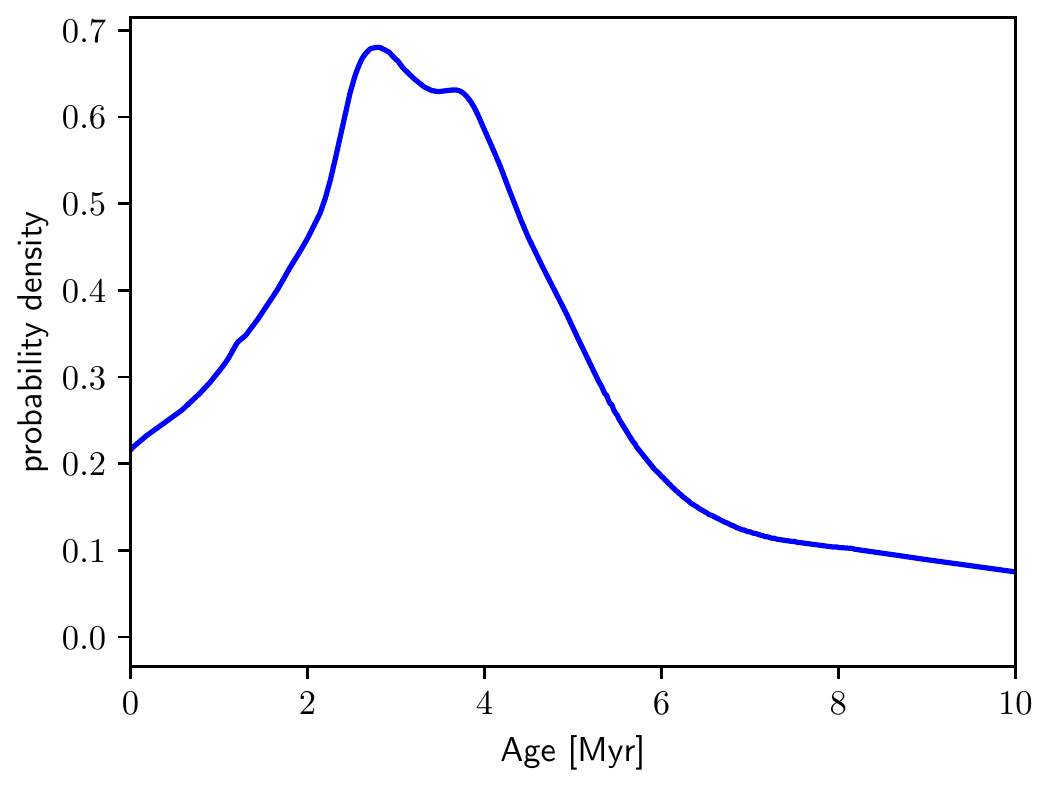} 
 \caption{Probability density functions of ages.}
   \label{fig5}
\end{center}
\end{figure}
This can be done as part of the analysis or in the post-processing, e.g. with BONNSAI (\cite[Schneider et al. 2014]{Schneider2014}), when determining stellar masses and ages for our sample. Figure~\ref{fig5} shows the age probability density 
of our sample based on ages calculated with BONNSAI on the bases of the stellar evolutionary tracks of \cite[Brott et al. (2011)]{Brott2011} and \cite[K\"ohler et al. (2015)]{Koehler2015}. The probability density is comparable to the findings of \cite[Schneider et al. (2018)]{Schneider2018}, but features are smoother and less prominent as a result of the overall larger uncertainties. However, taking into account that our analysis was completed in less then 2 weeks while the study of \cite[Schneider et al. (2018)]{Schneider2018} was based on 4 PhD theses with a combined effort of more than 10 years.

\section{Conclusion}

With the advent of large spectroscopic surveys such as WEAVE\footnote{{https://www.ing.iac.es/weave}} and 4MOST\footnote{{https://www.4most.eu}} 10\,000s of massive stars are going to be observed and need to be analysed. We will require fully automated spectroscopic analysis tools which reduce the human interaction to a minimum to cope with the amount of data. The pipeline presented here analyses $\sim$ 250 stars in less than half a day. Overall the quality of the spectroscopic fits is good, but around 15\% of the stars need additional attention as a result of strong nebular contamination, low S/N, multiplicity et cetera. Still this will provide more time for doing science rather than data analysis.

Future developments will replace the $\chi^2$-minimiser with machine learning approach including observational and model errors. After training the neural network the analysis process should be faster, less resource hungry and increasing the accuracy.

\section*{Acknowledgements}
JMB is supported by the Science and Technology Facilities Council research grant ST/V000853/1 (PI. V. Dhillon).

\begin{discussion}

\discuss{Fabry}{As you mentioned the $\log g$ uncertainty is driven by normalisation issues (Paschen lines) or uncertainties on the normalisation. Do you have suggestions, how to improve on this or will machine learning solve this?}

\discuss{Bestenlehner}{You can use machine learning to normalise your spectra, but if your analysis pipeline uses machine learning the normalisation is less of an issue as preliminary tests have shown. Here we use a simple normalisation routine to simulate the analysis work-flow of 4MOST. However, if spectra are flux calibrated and the temperature is reasonably well known, they can be normalised by dividing the observation by the theoretical model energy distribution (SED). In this way the Paschen jump is better normalised out and the accuracy of $\log g$ improves.}

\discuss{Najarro}{Large errors on logg comes from high Paschen series which are not well fitted. In comparison with VFTS, where the $\log g$ are based on low order Balmer series, $\log g$ on the Paschen series is unconstrained. }

\discuss{Bestenlehner}{The Paschen lines contain a lot of information on $\log g$, but the line profiles overlap and become not only very sensitive to the normalisation but also to the line broadening parameters (macro-turbulent and projected rotational velocities). In addition, the Paschen line profiles are less verified in stellar atmosphere codes than the Balmer series. The machine learning approach might improve on this in the future, but I would like to understand first the issues and how well my current approach works before using a new approach where I want to be able to follow step by step the analysis process like using a black-box.} 

\end{discussion}

\end{document}